\documentclass[twocolumn,showpacs,preprintnumbers,amsmath,amssymb]{revtex4}
\usepackage{graphicx}% Include figure files
\usepackage{graphics}% Include figure files
\usepackage{dcolumn}% Align table columns on decimal point
\usepackage{bm}% bold math
\begin{document}

\title{Coherent field emission image of graphene predicted with a microscopic theory}

\author{Zhibing Li\footnote{corresponding author: stslzb@mail.sysu.edu.cn} and Xingsheng Xu}

\address{The State Key Laboratory of Optoelectronic Materials and
Technologies \\ School of Physics and Engineering, Sun Yat-Sen
University, Guangzhou, P.R. China}

\author{H.J. Kreuzer}

\address{Department of Physics and Atmospheric Science, Dalhousie University Halifax NS B3H 3J5, Canada}

\begin{abstract}  Electrons in the mono-layer atomic sheet of graphene have a long coherence length
of the order of micrometers. We will show that this coherence is
transmitted into the vacuum via electric field assisted electron
emission from the graphene edge. The emission current density is given analytically. The parity of the carbon pi-electrons leads to an image
whose center is dark as a result of interference. A dragonfly pattern with a dark body perpendicular to the edge is predicted for the armchair edge whose emission current density is vanishing with the mixing angle of the pseudo-spin. The interference pattern may be observed up to
temperatures of thousand Kelvin as evidence of coherent
field emission.
Moreover, this phenomenon leads to a novel coherent electron line
source that can produce interference patterns of extended objects
with linear sizes comparable to the length of the graphene edge.
\end{abstract}


\maketitle

Nano-emitters of various shapes and materials have been used for
cold field electron emission (CFE) with the aim of constructing a
practical microelectronic vacuum electron source that may be used in
flat-panel displays, electron microscopes and parallel e-beam
lithography systems. The feature most important for such
applications is the high aspect ratio of nano-tips enhancing the
apex field to the point where field electron emission can be driven
by macroscopic electric fields of about ten volts per micrometer or
even less. Two successful examples of this approach are Spindt-type
cathodes\cite{spin68} and single atom field emission tips.\cite{fink88,fink91} In the context of
the present paper we should note that in the latter the emission
volume is the terminating atom so that emitted electrons are
spatially coherent as proven succinctly in in-line electron
holography.\cite{horc93} The coherent field emission of carbon-based
nano-emitters has been confirmed with the perfect interference
fringes produced by the electron beam from carbon nanotube CFE.\cite{oshi02,hass10}

Several groups have demonstrated that graphene does show promising
CFE properties such as a low emission threshold field and large
emission current density.\cite{eda08,lee09,male08,qian09,wu09,zhen09,xiao10} What we will show in this paper is
that the unique electronic properties of graphene and its monolayer
atomic structure result in a novel electron line source with
coherent field emission that can produce interference patterns of
extended objects with linear sizes comparable to the length of the
graphene edge. What makes graphene unique is that its
two-dimensional electron system is well described by a superposition
of 2p carbon orbitals and that its low energy excitations
behave like Dirac particles.\cite{novo05,neto09,das11} Due to the perfect atomic structure,
electrons in graphene have long coherent lengths (a few micrometers
at room temperatures \cite{geim09}). We will show that the emission electrons
will carry this coherence of the quantum states of graphene into the
vacuum.

The challenge is to explain/predict the coherent aspect of the CFE
image. Because of its wave nature coherent emission electrons must
have been in a collective quantum state in the solid and then be
transmitted through the tunneling barrier into a single particle
state in the vacuum. To address the coherence aspect of CFE, one
must incorporate the electronic structure of a
mesoscopic/macroscopic nano-emitter and deal with the screening of
the macroscopic applied field and of the atomic orbitals along the
edges so that one is in a position to calculate the tunneling wave
function. Needless to say this goes beyond the framework of
conventional CFE theory as the phase of the quantum states is
crucial here. But it can be done for graphene as shown in this paper. We
will construct the wave function of the emitted electron
analytically thus extracting important information about the
collective quantum state.

\begin{figure}
\includegraphics{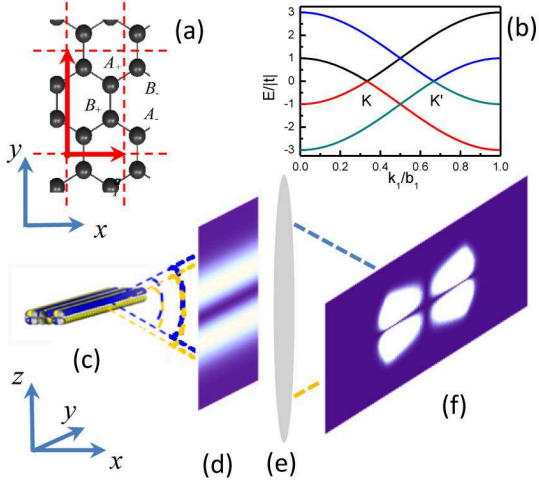}% Here is how to import EPS art
\caption{\label{fig:1} (a) A piece of graphene with the
black balls representing the carbons and the bond length $a =0.142$ nm. The armchair edge is in the $y$-direction. The red rectangular plaquette is a unit cell with the
unit vectors ${\bf a}_{1} =(\sqrt{3}a, 0)$ and ${\bf a}_{2} =(0, 3a)$. (b) The energy bands of the Brillouin zone. The unit
vectors of the reciprocal lattice are ${\bf b}_{1}=(2\pi/(\sqrt{3}a),0)$ and ${\bf b}_{2}=(0,
2\pi/(3a))$. (c) The wave function profile of the HOMO along the armchair edge (DFT simulation). (d) The emission image near the anode plane in a field $F_{0}$=20V/$\mu$m at the room temperature. (e) The electric objective lens. (f) The image on the back focal plane of the objective lens.}
\end{figure}

A piece of graphene with the armchair edge (AE) along the $y$-axis is shown in Fig.\ref{fig:1}(a). If it rotates by an angle of 90 degrees on the
plane, the emission edge will become the zigzag edge (ZE). We adopt
a rectangular plaquette of the red lattice as the unit cell (see Fig.\ref{fig:1}(a)). There are four carbons in a unit cell, each is labeled by $\alpha=A_{\pm}$ or $B_{\pm}$
and referred to as the $\alpha$-carbons. The energy bands are plotted in Fig.\ref{fig:1}(b).
The CFE experiment is schematically shown in Fig.\ref{fig:1} (c-f), with (c) the graphene, (d) the anode plane, (e) the electric objective lens and (f) the back focal plane(BFP), respectively. A uniform
macroscopic field $F_{0}$ is applied in the negative direction of the $x$%
-axis.

The basic microscopic process of the CFE may be understood as an
electron in the graphene being annihilated and then
created in the out-going state in the vacuum. This is the transition of a two-dimensional massless
exciton to a three-dimensional elementary electron.
The low energy excitations obey Dirac dispersion relations $E({\bf k})=\pm \hbar v_{F}|{\bf k}|$ , with
$\pm$ for the positive/negative
energy bands, $v_{F}\sim c/330$
the Fermi velocity,\cite{neto09} and ${\bf k}$ a wave vector relative to the $K$ point (Fig.\ref{fig:1}(b)) on the $x$-$y$ plane. We will focus on the positive energy bands which has dominate contribution to the CFE, although our analysis can be easily extended to including the negative energy bands.
The eigenstates are Bloch waves along the $y$-direction but not in the $x$-direction due to the edge. We have adopted the time-Y parity representation 
such that the eigenstates of the half-infinite graphene have either even ($s=+$) or odd ($s=-$) parity in the combination of the
time-reversal and the $y$-directional reflection transformations. The eigenfunctions can be expanded in the Wannier-Bloch hybrid orbitals that in the $x$-direction are localized at each carbon and extended in the $y$-direction,
\begin{equation}
\psi_{{\bf k}s}({\bf r})=\sum_{m,\alpha,l}b_{2} C^{\alpha}_{s,m}({\bf k})\varphi_{m,{\bf k}l}(x-x^{\alpha}_{m},z)\frac{e^{ik_{y}^{l}y}}{\sqrt{2\pi}}
\label{eigenstate}
\end{equation}
where $b_{2}=\frac{2\pi}{3a}$ is the reciprocal lattice spacing in the $y$-direction, $k_{y}^{l}=k_{y}-lb_{2}$ with $l$ an integer, and $\varphi_{m,{\bf k}l}(x,z)$ is defined as the $k_{y}^{l}$-component of the $y$-directional Fourier transformation of the Wannier orbital. The Bloch-Wannier orbital with $k_{y}^{l}$ has $y$-directional kinetic energy $E_{y}=\frac{(\hbar k_{y}^{l})^2}{2m}$ and the energy for the motion on the $x$-$z$ plane is $E_{p}=E({\bf k})-E_{y}$. This leads to an effective vacuum barrier height $W_{p}=W_{o}-E_{p}$, with $W_{o}\sim 4.32$ eV the work function of graphene,\cite{wang11a} for tunneling normal to the edge.\cite{tada02,huan05}  For hydrogen saturating edges,
$E_{y}$ of the AE vanishes at the $K$ point if $l=0$ while it is 2.8eV for the ZE.
Therefore emission from AE with $k_{y}^{0}=k_{y}$ is dominant. However we should note that
the transverse wave numbers of edge states of the ZE are shifted to the
vicinity of the $\Gamma$ point if the edge is saturated via atoms with
p-orbitals (oxygen for instance) with local energy similar to the 2p orbital of carbon.\cite{klei94}
In such a case CFE from edge states of ZE is also possible.
In the present paper we will focus on the band state emission and only consider the $l=0$ mode of the AE. The subscript $l$ will be dropped latter on.

The amplitudes $C^{\alpha}_{s,m}({\bf k})$ are determined by the boundary condition and the time-Y parity.
Notably, the AE will mix states with the same bulk energy and a given $k_{y}$. These degenerate states include excitations from both Dirac cones ($K$ and $K^{\prime}$ in Fig.\ref{fig:1}(b)) and states with $k_{x}$ and $-k_{x}$. In the final expression for the emission current density, we only need the combination $\widetilde{C}_{{\bf k}}=|C^{A_{+}}_{s,1}({\bf k})+C^{B_{+}}_{s,1}({\bf k})|=\frac{\sin\frac{\pi}{3}|\sin\gamma|}{\sqrt{MN}}$
with $\gamma$ the pseudo-spin mixing angle (the phase difference between the two sublattices) given by $\tan\gamma=k_{y}/k_{x}$,\cite{neto09} and $M$ ($N$)  the total number of rows (columns).

Because the relevant electronic properties of graphene can be attributed to the 2p orbital of carbon, the edge Wannier orbital should have the form $\varphi_{{\bf k}}({\bf \rho})=N_{{\bf k}}z\sqrt{\varrho} e^{-\tau_{p} \rho}$ with ${\bf \rho}=(x,z)$ and $\varrho=\sqrt{x^2+z^2}$ inside the region where the atomic potential is dominant (APD region). Here the factor $z$ comes from the 2p orbitals, $N_{{\bf k}}$ is a normalization factor and $\tau_{p}=\frac{1}{\hbar}\sqrt{2mW_{p}}$.

The graphene eigenstates are not exactly stable in the applied fields. The emission wave should be found by solving the Schr{\"o}dinger equation. The probability amplitude emitted from $\psi_{{\bf k}s}$ for large times is strictly given by the path-decomposition formula,\cite{auer84}
\begin{equation}
\psi^{em}_{{\bf k}}({\bf r})=\int_{\Sigma}G^{r}({\bf r},{\bf r}^{\prime})\frac{i\hbar \overleftrightarrow{\partial_{n}}}{2m} \int_{{\bf r}^{\prime\prime}} G({\bf r}^{\prime},{\bf r}^{\prime\prime})\psi_{{\bf k}s}({\bf r}^{\prime\prime})
\label{emissionwave}
\end{equation}
where $\int_{\Sigma}$ is the surface integral over an arbitrary surface $\Sigma$ that separates the graphene and the vacuum, and $\overleftrightarrow{\partial_{n}}$ is the right normal derivative minus the left one on $\Sigma$. The $G$ is the retarded fixed energy Green's function and the restricted Green's function $G^{r}$ is defined by a path integral over energy-fixed paths that stay strictly outside $\Sigma$ and do not touch it except at the initial point. The physical meaning of (\ref{emissionwave}) is clear. The emission electron must go
through the separation surface $\Sigma$ on which the amplitude $G\psi_{{\bf k}s}$  at each
point acts as a point source. The operator $-\frac{\hbar}{2m}\overleftrightarrow{\partial_{n}}$
represents the mean imaginary flux through the surface. The
Green's function $G^{r}$  propagates the wave from $\Sigma$ to the screen. Thus the surface
integral gives the emission wave as the coherent superposition of
waves from all points on $\Sigma$. We will choose $\Sigma$ as a surface of constant potential in the vacuum barrier region such that only the least action path has significant contribution to $G^{r}$ and the semi-classical approximation (JWKB) is applicable. We also require that $\Sigma$ locates on the boundary of the ADP region so $\Sigma$ is invariant in the $y$-direction yet the orbital wave function is still feasible on it. Because $G({\bf r}^{\prime},{\bf r}^{\prime\prime})$ decreases with $|{\bf r}^{\prime}-{\bf r}^{\prime\prime}|$ exponentially, only the edge carbon orbitals with $m=1$ in the expansions (\ref{eigenstate}) have significant contribution to (\ref{emissionwave}).

To go further, the vacuum
barrier potential in the vicinity of the edge is needed. It may be written
as $V({\mathbf \rho})=W_{o}+V_{at}({\mathbf \rho})+V_{e}({\mathbf \rho})$
with $V_{at}$ the short-ranged atomic potential and $V_{e}({\mathbf \rho})$  the electric
potential energy.
First note that the $y$-directional momentum $\hbar k_{y}$ is conserved due to the symmetry.
More complicate interactions would be involved when the electron is very close to the atom core. But the exact form of $V_{at}$ close to the atomic core is not
crucial if the wave function is known in the APD region as we have assumed. A feasible model is $V_{at}=-\frac{P x}{\varrho^2}$ with $P$ the
dipole line density along the edge.\cite{wang11a} It is easy to obtain $V_{e}$ outside the APD region (referred to as the off-APD region) via the conformal transformation, $\widetilde{z}+i\widetilde{x}=\sqrt{(z+i(x+h))^2+h^2}$ with $h$ the height of the graphene. On the $\widetilde{{\bf \rho}}=(\widetilde{x},\widetilde{z})$ plane,
$V_{e}=-eF_{0}\widetilde{x}$. With the screen far away from the graphene, the observed image is produced by the near axis ($x$ or $\widetilde{x}$-axis) electrons in the off-APD region. In the near axis approximation (NAA), neglecting terms of order $\frac{\widetilde{z}^{2}}{\widetilde{x}^2}$, one gets $V_{at}\sim -\frac{2hP}{\widetilde{x}^2}$ and the kinetic energy operator for the vacuum electron $\hat{K}=-\frac{\hbar^2}{2m}\frac{h^2+\widetilde{x}^2}{\widetilde{x}^2}(\partial^{2}_{\widetilde{x}}+\partial^{2}_{\widetilde{z}})$.

In the coordinates $(\widetilde{x},y,\widetilde{z})$ $\Sigma$ of (\ref{emissionwave}) becomes a plane with a constant $\widetilde{x}$. We choose the separation point at the maximum of $V(\widetilde{x})$, which is $\widetilde{x}_{p}=\frac{2W_{p}}{3eF_{0}}$. Thus it is easy to obtain $\psi^{em}_{{\bf k}}$ using the JWKB approximation for the Green's functions.
We assume the potential in the region beyond the anode to be a constant $-V_{a}$. The emission wave arriving on the anode plane
is the analogue of the two-slit Young's interference with an
effective slit constant $\widetilde{z}_{p}=\sqrt{\frac{h}{\tau_{p}}}$ and an effective wave number $k_{a}=\frac{1}{\hbar}\sqrt{2mV_{a}}$. For a large focal length $x_{f}$ and small emission angles, the wave function on the BFP is the Fourier transformation of the wave function on the anode plane that turns out to be
\begin{equation}
\psi^{em}_{{\bf k}}({\bf r}_{f})=c_{0}\frac{\widetilde{C}_{{\bf k}}\tau_{p}^{\frac{1}{2}}S_{p}^{\frac{5}{4}}}{a x_{f}}e^{-\frac{b_{0}}{2}S_{p}}\phi_{p}(\theta_{z})\delta(k_{y}-\theta_{y}k_{a})
\end{equation}
where $\theta_{y}=\frac{y_{f}}{x_{f}}$ and $\theta_{z}=\frac{z_{f}}{x_{f}}$ are emission angles with respect to the $y$ and $z$ axes respetivelty,  $S_{p}=(\frac{\widetilde{x}_{p}}{\widetilde{z}_{p}})^{2}$ and $\phi_{p}(\theta_{z})=(\frac{2 \mu_{p}^{3}}{\sqrt{\pi}})^{\frac{1}{2}}\theta_{z}e^{-\frac{\mu_{p}^{2}}{2}\theta_{z}^{2}}$ with $\mu_{p}=(1+\log(2+\sqrt{3}))^{\frac{1}{2}}k_{a}\widetilde{z}_{p}$. The factors $c_{0}\approx 0.890$ and $b_{0}=1+\frac{4}{5\sqrt{3}}$ for the present $\widetilde{x}_{p}$. Their values will change with $\widetilde{x}_{p}$ but the effect will be compensated by the change of $S_{p}$ and the wave function is only slightly affected.

\begin{figure}
\includegraphics{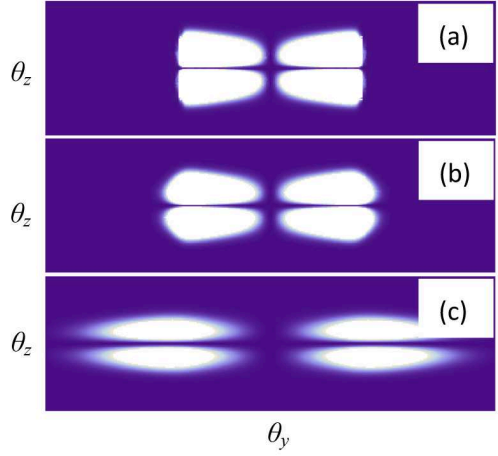}% Here is how to import EPS art
\caption{\label{fig:2} (Color online) The image on the back focal plane for a graphene height $h=10\mu$m and an anode distance $x_{a}=1$mm in the field $F_{0}=20V/\mu$m. The showing region has $|\theta_{y}|<0.005$ and $|\theta_{z}|<0.00015$. The panels (a), (b), and (c) are for temperatures 100K, 300K, and 1000K, respectively.
}
\end{figure}

With the emission wave function obtained, it is straight forward to find the emission current density of each individual graphene state. The total emission current is a non-coherent summation of the individual currents weighted by the Fermi-Dirac distribution(FDD). Because the emission depends strongly on the barrier height and is
also controlled by the FDD, it is
expected that only modes very close to the Fermi energy ($E_{F}$) can be
emitted. This is the reason why CFE leads to observable interference even at high temperatures. Therefore the value of $E_{F}$ is crucial for the emission current and should be determined taking account also of the induced excess charge density. Using the density of states of the tight-binding model and an iteration method we find $E_{F}=\hbar v_{F}\sqrt{\frac{4\pi \varepsilon_{0}F_{0}\sqrt{h}}{e\sqrt{2 \sqrt{3}a}}}$ at the edge, consistent with a recent estimate.\cite{wang11b}
The angular distribution of the emission current on the BFP now reads
\begin{eqnarray}
f(\theta_{y},\theta_{z})&=&\frac{3^{\frac{7}{2}}c_{0}^{2}}{(2\pi)^{3}}\frac{aNek_{a}}{2mv_{F}}\int^{\infty}_{E_{a}|\theta_{y}|} dE \frac{E\sin^{2}\gamma}{\sqrt{E^2-E_{a}^{2}\theta_{y}^{2}}} \nonumber \\
&&\cdot \frac{e^{-b_{0}S_{p}}}{e^{\beta(E-E_{F})}+1}\phi_{p}(\theta_{z})^{2}
\label{jcd}
\end{eqnarray}
where
$E_{a}=\hbar v_{F} k_{a}$. Because $\sin\gamma=\frac{E_{a}}{E}\theta_{y}\sim \theta_{y}$ and $\phi_{p}\sim \theta_{z}$,
it is clear that the current intensity is
zero and the BFP image is dark in the center cross with $\theta_{y}\theta_{z}=0$. The pattern shown in Fig.\ref{fig:1}(f) (and more in Fig.\ref{fig:2}) is like a dragonfly. The dark slit that parallel to the graphene edge is caused by the destructive interference of two partial waves emitted from the
p-orbital that has two lobes with the phase difference $\pi$ on either side of the graphene plane (Fig.\ref{fig:1}(c)). Because $\phi_{p}^{2}$ only weakly depends on $E({\bf k})$ and $\theta_{y}$, one may move it out off the integral (\ref{jcd}) and see that the brightest point of a wing has an angle to the $y$-axis of $\mu_{F}^{-1}$ in which $E_{p}$ is substituted with $E_{F}$ as the subscript $F$ indicated. Clearly this angle is almost independent of the temperature. In Fig.\ref{fig:2}, the image for $T=100$ (a), $300$K (b), and $1000$K (c) are presented. In the low temperatures, the wing tips are sharply cut at $\theta_{y}\sim \pm \frac{E_{F}}{E_{a}}$ . In the high temperatures, the wing tips are smeared while the side-edges are still sharp. On the other hand, the dark ``dragonfly" body has different originator. It is because that the angular distribution is proportional to $\sin^{2}\gamma$ while $\gamma$, the mixing angle of the pseudo spin, is vanishing for the eigenstates with $k_{2}=0$. Before the lens where the waves of different $k_{2}$ have not been dispersed, one will see two bright bands parallel to the $y$-axis(Fig.\ref{fig:1}(d)).

\begin{figure}
\includegraphics{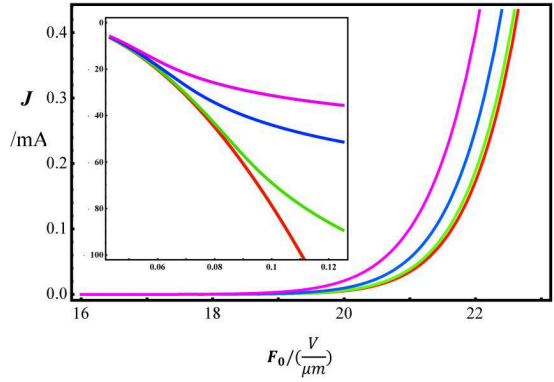}% Here is how to import EPS art
\caption{\label{fig:3} (Color online) The total emission currents versus the applied fields for $h=10\mu$m and an edge length $3aN=1$mm. The curves from bottom to top are the total currents for temperatures 100K, 300K, 600K, and 900K respectively. The inset shows the corresponding Fowler-Nordheim plots with $\frac{1}{F_{0}}$/($\frac{\mu m}{V}$) as the horizontal axis and $\log(JF_{0}^{-2})$/($\log(mA(\mu m)^{2}V^{-2})$) as the vertical axis.}
\end{figure}

The total emission current $J$ may be given by the integration of (\ref{jcd}). But it is more convenient to calculate it with the wave function on the anode plane. The $J$-$F_{0}$ characteristic is presented in Fig.\ref{fig:3} where the four curves from bottom to top are total currents for the temperatures $100$, $300$, $600$, and $900$K, respectively. The corresponding Fowler-Nordheim plots are also given as the inset of Fig.\ref{fig:3}.
In the CFE regime one may expand the emission energy around $E_{F}$ and obtain an explicit expression for the total emission current,
\begin{eqnarray}
J&=&\frac{c_{0}^{\prime}aNeE_{F}(W_{F}S_{F})^{\frac{3}{2}}}{m^{\frac{1}{2}}(\hbar v_{F})^{2}}e^{-b_{0}S_{F}} \nonumber \\
& &\cdot \frac{\alpha_{T}}{\sin\alpha_{T}}[I_{0}(\alpha_{F}E_{F}^{2})-I_{1}(\alpha_{F}E_{F}^{2})]e^{-\alpha_{F}E_{F}^{2}}
\label{Jtot}
\end{eqnarray}
where $c_{0}^{\prime} \approx 0.092$, $\alpha_{T}=\frac{5\pi b_{0}S_{F}}{2\beta W_{F}}$ and $\alpha_{F}=\frac{5\pi b_{0}S_{F}}{8mv_{F}W_{F}}$ with the subscript $F$ indicating the substitution of $E_{p}$ by $E_{F}$ in the corresponding quantities, and $I_{0}$ ($I_{1}$) is the modified Bessel functions of the zero (first) order respectively. For the low temperatures and low fields such that both $\alpha_{T}$ and $\alpha_{F}E_{F}^{2}$ are small, the second line of (\ref{Jtot}) is not sensitive to either the temperatures or fields. The characteristic of the total emission current is mainly determined by the factor $e^{-b_{0}S_{F}}$ that is only slightly different from the conventional CFE theory.\cite{qin11} Note that, as a characteristic feature of the two-dimensional CFE, $J$ depends on $h$ and $F_{0}$ through $S_{F}$ and $E_{F}$ in a combination of $F_{0}\sqrt{h}$.

In summary, we have presented an analytical treatment of field emission from
graphene at the microscopic level, including analytical expressions
for the vacuum barrier, the Fermi energy, and the emission wave
function in three-dimensional space. The Fermi energy is field depending due to the field penetration.
The dragonfly-like
emission pattern is predicted.
The origin of the dragonfly pattern is different from the interference of
coherent beams from different atoms or atom clusters that give a bright center, but is
associated to the symmetry of the p-orbital of carbon and the bulk states of the graphene. The dark slit in the center of the pattern and parallel to the edge is a result of the parity of the
p-orbital of the carbon atoms. We have proved that the width of the dark slit is almost independent of the temperatures. An less expected fact is the dark body of the ``dragonfly" that originates from the graphene structure.
Intuitively, the
uniform edge of graphene plays the role of a one-dimensional
electron grating. Along the edge of graphene, the
transverse momentum of the emission is equal to the Bloch momentum
of the graphene. The
vacuum barrier allows emission only from states within a narrow
range of energy close to the Fermi energy and with small transverse
kinetic energy. Because the mixing of electron states of the two sublattices, the emission of vanishing $y$-directional kinetic energy is absent, resulting the dark ``dragonfly" body. Interestingly, the wing tips turn out to be brightest region. That implies the transverse kinetic energy is maximized for a given total kinetic energy controlled by $E_{F}$. The appearance paradox may be solved by the fact that the massless exciton of the graphene acquires a nonzero mass in vacuum. It is the massive real electron that tunnels through the vacuum barrier and its transverse kinetic energy is suppressed by the mass. Last but not least, the dark slit between a pair of wings on either side of the pattern may be smeared out by a magnetic flux parallel to the edge but the dark ``dragonfly" body can not be changed.

Our theory may get support from a recent experiment of Yamaguchi et al \cite{yama11} who show two field emission patterns
from a graphene edge in the configuration considered by us. At a
voltage of 5.1 V they see a regular interference pattern with the
center the brightest that they attribute to interference between
regular patches of oxygen atoms adsorbed on the edge. However, at a
lower emission voltage of 2.1 V they see indeed a pattern that is
dark in the forward direction flanked by two bright emission cones
as predicted by our theory. The unusual up bending Fowler-Nordmeim plot (the inset of Fig.\ref{fig:3}) in the regime of weak fields and/or short graphene height has been confirmed by another experiment. \cite{xiao10}

Acknowledgment: The project was supported by the National Basic
Research Program of China (2007CB935500 and 2008AA03A314). HJK
acknowledges support by the Natural Sciences and Engineering Council
of Canada and the Office of Naval Research, Washington DC.

\clearpage

\end{document}